\documentclass[aps, twocolumn, superscriptaddress, showpacs, nofootinbib, longbibliography]{revtex4-1}

\usepackage[utf8]{inputenc}
\usepackage[T1]{fontenc}
\usepackage{ae,aecompl} 
\usepackage{graphicx}
\usepackage{amsmath}
\usepackage{color}
\usepackage{amssymb}
\usepackage{latexsym}
\usepackage{wasysym}
\usepackage{psfrag}
\usepackage{ifthen}
\usepackage[citecolor=blue,colorlinks=true]{hyperref}
\usepackage{longtable}
\usepackage{float}
\usepackage[utf8]{inputenc}
\usepackage{lineno}

\usepackage{aas_macros}
\usepackage[normalem]{ulem}

\definecolor{bbsalmon}{rgb}{1.0, 0.47, 0.42}
\definecolor{datablue}{rgb}{0.0, 0.0, 1.0}

\begin{document}

\preprint{APS/123-QED}

\title{Prospects for detecting and localizing short-duration transient gravitational waves from glitching neutron stars without electromagnetic counterparts}%

\author{Dixeena Lopez}
\author{Shubhanshu Tiwari}%
\affiliation{%
 Physik-Institut, University of Zurich, Winterthurerstrasse 190, 8057 Zurich, Switzerland 
}%

\author{Marco Drago}%
\affiliation{Dipartimento di Fisica, Università di Roma ``La 
 Sapienza'', Piazzale Aldo Moro 2, I-00185 Roma, Italy
 }%
  \affiliation{INFN, Sezione di Roma, Piazzale Aldo Moro 2, I-00185 Roma, Italy
 }%

\author{David Keitel}%
 \affiliation{Departament de  F\'isica, Institut d’Aplicacions Computacionals i de Codi Comunitari (IAC3), Universitat de les Illes Balears, and Institut d’Estudis Espacials de Catalunya (IEEC), Carretera de Valldemossa km 7.5, E-07122 Palma, Spain}%

\author{Claudia Lazzaro}%
\affiliation{%
Dipartimento di Fisica ed Astronomia, Università degli Studi di Padova, Via G. Marzolo, 8 - 35131 Padova, Italy}%
\affiliation{INFN, Sezione di Padova, I-35131 Padova, Italy} 

\author{Giovanni Andrea Prodi}%
\affiliation{University of Trento, Physics Department and INFN,
Trento Institute for Fundamental Physics and Applications, via Sommarive 14, 38123 Povo, Trento, Italy}

\begin{abstract}


Neutron stars are known to show accelerated spin-up of their rotational frequency called a glitch. Highly magnetized rotating neutron stars (pulsars) are frequently observed by radio telescopes (and in other frequencies), where the glitch is observed as irregular arrival times of pulses which are otherwise very regular. A glitch in an isolated neutron star can excite the fundamental (\textit{f})-mode oscillations which can lead to gravitational wave generation. This gravitational wave signal associated with stellar fluid oscillations has a damping time of $10-200$\,ms and occurs at the frequency range between $2.2-2.8$\,kHz for the equation of state and mass range considered in this work, which is within the detectable range of the current generation of ground-based detectors. Electromagnetic observations of pulsars (and hence pulsar glitches) require the pulsar to be oriented so that the jet is pointed toward the detector, but this is not a requirement for gravitational wave emission which is more isotropic and not jetlike. Hence, gravitational wave observations have the potential to uncover nearby neutron stars where the jet is not pointed towards the Earth. In this work, we study the prospects of finding glitching neutron stars using a generic all-sky search for short-duration gravitational wave transients. The analysis covers the high-frequency range from $1-4$\,kHz of LIGO--Virgo detectors for signals up to a few seconds. We set upper limits for the third observing run of the LIGO--Virgo detectors and present the prospects for upcoming observing runs of LIGO, Virgo, KAGRA, and LIGO India. We find the detectable glitch size will be around $10^{-5}$\,Hz for the fifth observing run for pulsars with spin frequencies and distances comparable to the Vela pulsar. We also present the prospects of localizing the direction in the sky of these sources with gravitational waves alone, which can facilitate electromagnetic follow-up. We find that for the five detector configuration, the localization capability for a glitch size of $10^{-5}$\,Hz is around $132\,\mathrm{deg}^{2}$ at $1\sigma$ confidence for $50\%$ of events with distance and spin frequency as that of Vela. 

\end{abstract}

\maketitle

\section{Introduction}

Neutron stars (NSs) are one of the most promising and versatile sources of gravitational waves (GWs)~\cite{fullGR_Glampedakis},
including both isolated NSs and those in binary systems with other compact objects.
Several searches use varied methods for different scenarios depending on the nature of the targeted GW signals.
Advanced LIGO \cite{ligo_2015} and Advanced Virgo \cite{virgo_2014} have detected GW signals from compact binary coalescences (CBCs), including binary neutron star coalescences and neutron star--black hole coalescences \cite{gwtc1,gwtc2,gwtc3}.
Nonradial oscillation modes, magnetic or thermal mountains for both isolated NSs, and those in binaries, as well as accretion in binary systems are among the sources for continuous GWs~\cite{Riles_2022}.
Isolated NSs are also an interesting astrophysical source for transient GWs in the detectable range of current generation GW detectors.
For example, searches have been conducted for magnetars that can be strong emitters of transient GWs and short bursts of $\gamma$ rays \cite{magnetar_2019,GRBO3}, but no detection has been made yet. 

In this paper, we focus on transient GWs from glitching pulsars.
Rotating isolated NSs, including pulsars, generally show a decrease in their spin frequency over time. However, some exhibit a sudden jump in their rotation frequency known as glitches \cite{Fuentes:2017bjx}.
So far, at least 740 glitches from 225 known pulsars have been reported with glitch sizes of $\Delta\nu_{s}\approx 10^{-9}$--$10^{-4}$\,Hz~\cite{gbank_2019,315_glitches, glitch_stat, glitch_bank,atnf_bank,Moragues:2022aaf}.

Glitches in isolated neutron stars can excite acoustic and inertial stellar oscillations which in turn generate GWs lasting  $\lesssim 0.2$\,s at frequencies from 1--3\,kHz depending on the models and source parameters. The $f$-mode oscillations are among these potential causes of GW emission \cite{Andersson1998TowardsGW,WH}. Recently, a scenario for GWs from $f$ modes in smaller glitch candidate events was also studied \cite{Yim_2022}.
Historically, a first targeted search for short transient GWs associated with a glitch was conducted for a Vela pulsar glitch in August 2006, finding no evidence of GWs \cite{S5_NS}.
More recently, a generic all-sky search for GW transients during the third observing run~\cite{allskyo3} was also interpreted under the glitch scenario, providing a limit on the minimum detectable glitch size around $10^{-4}$\,Hz  for an optimally oriented source and with Vela reference parameters. During the second observing run (O2) of Advanced LIGO and Advanced Virgo (November 2016--August 2017), four pulsar glitches were observed in radio telescopes based on \cite{glitch_bank}. Considering the whole glitch energy transformed as GW by the $f$-mode oscillation of NS, only the peak amplitude from the Vela glitch detected by the radio telescope on 2016 December 12 is above the power spectral density (PSD) of LIGO and Virgo detectors during the O2 run \cite{Palfreyman_vela,vela_2016}. During the glitch in Vela, only LIGO Hanford was online out of three GW detectors, making the GW counterpart detection unreliable \cite{2016_hanford}.
In addition, searches for longer-duration quasimonochromatic transient GWs correlated to pulsar glitches during the second and third observing runs \cite{long-duration,Narrowband_2021,Modafferi:2021hlm} also put upper limits on GW strain under that scenario. Moreover, a recent study about the prospects for observing longer-duration GW signals with current and future ground-based detectors is given in \cite{Moragues:2022aaf}.
However, in this paper, we focus on shorter signals from $f$ modes.

In general, the population of isolated NSs observed by electromagnetic (EM) observatories is a small fraction of the actual NS population in our Galaxy.
Hence, all-sky GW searches have the potential to find previously undiscovered NSs.
Follow-up searches of GW detections by EM observation, e.g., in the X-ray and radio bands,
could then help in constraining NS properties.
The sky localization information from the GW search is crucial to provide an opportunity for a targeted follow-up by EM telescopes.

This paper presents the all-sky search results for short-duration transient GWs from NS glitches during the third LIGO--Virgo observing run for arbitrarily oriented sources. We provide the prospects for future runs of the current generation of GW detectors regarding the glitch size one can probe. The future observing runs are expected to include KAGRA \cite{kagra}, LIGO India \cite{ligo_india},  and further upgrades of Advanced LIGO and Advanced Virgo. We also present the prospects for the sky localization of these sources for the upcoming observing runs. 

The paper is organized as follows. Section \ref{sec:signal} describes the signal model. Section \ref{sec:searches} discusses the search for short transient GW signals from glitching NSs. Section \ref{sec:localization} discusses the prospects of observing and localizing these GW signals for future ground-based detector searches. Section \ref{sec:discussions} discusses the results.

\section{Signal Model}
\label{sec:signal}

Two main mechanisms are considered to be responsible for pulsar glitches: starquakes and superfluid--crust interactions \cite{S5_NS,Clark_2007,models_glitch,gltich_mech,dynamics_glitch}. The energy generated during the excitation of oscillation modes by the starquakes or superfluid--crust interactions is given as \cite{WH,S5_NS,Clark_2007,Prix_2011}

\begin{equation} 
    \Delta E_{\mathrm{glitch}} \approx 4\pi^{2}I\nu_{\mathrm{s}}\Delta\nu_{\mathrm{s}} \,,
\end{equation}

where $I \sim 10^{38}$\,kg $\mathrm{m}^2$ is the stellar moment of inertia, $\nu_{\mathrm{s}}$ the spin frequency, and 
$\Delta \nu_{\mathrm{s}}$ the  increase in spin frequency \cite{S5_NS}. An order of magnitude estimate can be obtained by comparing with fiducial values of the frequency and its change during the glitch. The energy can be expressed as \cite{WH,S5_NS}

\begin{equation} 
    \Delta E_{\mathrm{glitch}} \approx 3.95 \times 10^{40} \mathrm{erg}\left(\frac{\nu_{\mathrm{s}}}{10\,\mathrm{Hz}}\right)\left(\frac{\Delta \nu_{\mathrm{s}}}{10^{-7}\,\mathrm{Hz}}\right) \,.
\end{equation}

A possible consequence of NS glitches is the excitation of one or more oscillations in the NS. This leads to the excitation of different families of pulsation modes like pressure $p$ modes (the fundamental of which is known as the $f$ mode) and the gravity $g$ modes corresponding to the energy of the glitch \cite{cowling_1941,Qnm_Kokkotas_99,WH, Superfluidity}. In this work, we are interested in the $f$ modes, which are the dominant mode in producing transient GWs from NS glitches  \cite{fg_mode1,fg_mode2,fg_mode3}.
For a perfectly spherical NS (nonrotating, nonmagnetic), the damping time and mode frequency are degenerate for each mode. Moreover, we consider only the dominant quadrupolar emission ($l=2$) here as higher-order modes ($l>2$) will be subdominant \cite{kip_dcc,kip} and also will occur at a higher frequency where the detectors lose sensitivity \cite{modes_reviews}. Hence, GWs associated with the excitation of pulsation modes ($f$ modes) are short-lived signals, which can be expressed in the time domain as \cite{WH,Kokkotas:2001ze}

\begin{equation}\label{damp_sinsoid} 
 h(t)= h_0 e^{-t/ \tau_\mathrm{gw} } \sin (2\pi\nu_\mathrm{gw}t) \,.
\end{equation}

Here, $h_0$ is the initial amplitude of the signal.
\mbox{$\nu_\mathrm{gw}$} and $\tau_\mathrm{gw}$ are the frequency and characteristic damping time of the signal, respectively.
 
The initial amplitude is related to the total GW energy emitted by a source at a distance $d$ \cite{WH,gw_luminosity98,S5_NS},
\begin{equation}  
    h_{0} = \frac{1}{\pi d \nu_\mathrm{gw} }\left(\frac{5G}{c^3}\frac{E_\mathrm{gw}}{\tau_\mathrm{gw}}\right)^{1/2} \,.
\end{equation}

Therefore, the peak GW amplitude of the $f$ mode ringdown signal, assuming the total energy generated by the excitation of the dominant mode is emitted as GWs ($E_\mathrm{gw}\approx E_{\mathrm{glitch}}$), is given as (Eq. (5) of \cite{WH}), 

\begin{equation} \label{eq_h0}  
\begin{aligned}
h_{0} =& \; 7.21 \times 10^{-24}\left(\frac{1\,\mathrm{kpc}}{d}\right)\left(\frac{\nu_{\mathrm{s}}}{10\,\mathrm{Hz}}\right)^{1/2}\\
&\left(\frac{\Delta \nu_{\mathrm{s}}}{10^{-7}\,\mathrm{Hz}}\right)^{1/2} \left(\frac{1\,\mathrm{kHz}}{\nu_\mathrm{gw}}\right)\left(\frac{0.1\,\mathrm{s}}{\tau_\mathrm{gw}}\right)^{1/2} \,.
\end{aligned}
\end{equation}

Quasinormal modes are classified according to the restoring force, which brings the perturbed element of the fluid back to the equilibrium position. We consider the nonrotating limit where only GWs from $f$-mode oscillations are well inside the sensitive bandwidth of ground-based GW detectors \cite{Andersson1998TowardsGW}.
Empirical relations for frequencies and damping times as functions of the mean mass density and compactness for various equations of state (EoS) are given as \cite{asteroseismology}

\begin{equation}\label{f_freq} 
 \nu_\mathrm{gw}[\mathrm{kHz}]=1.562+1.151\left(\frac{\bar{M}}{\bar{R}^3}\right)^{1 / 2}
\end{equation}
and
\begin{equation}\label{f_tau} 
\frac{1}{\tau_\mathrm{gw}[\mathrm{~s}]}=\frac{\bar{M}^3}{\bar{R}^4}\left[78.55-46.71\left(\frac{\bar{M}}{\bar{R}}\right)\right] \,,
\end{equation}
where
\begin{equation}
\bar{M}=\frac{M}{1.4\, M_{\odot}} \quad \text { and } \quad \bar{R}=\frac{R}{10\,\mathrm{km}} \,.
\end{equation}

\begin{figure}[t!] 
  \centering
    \scalebox{0.30}{\includegraphics{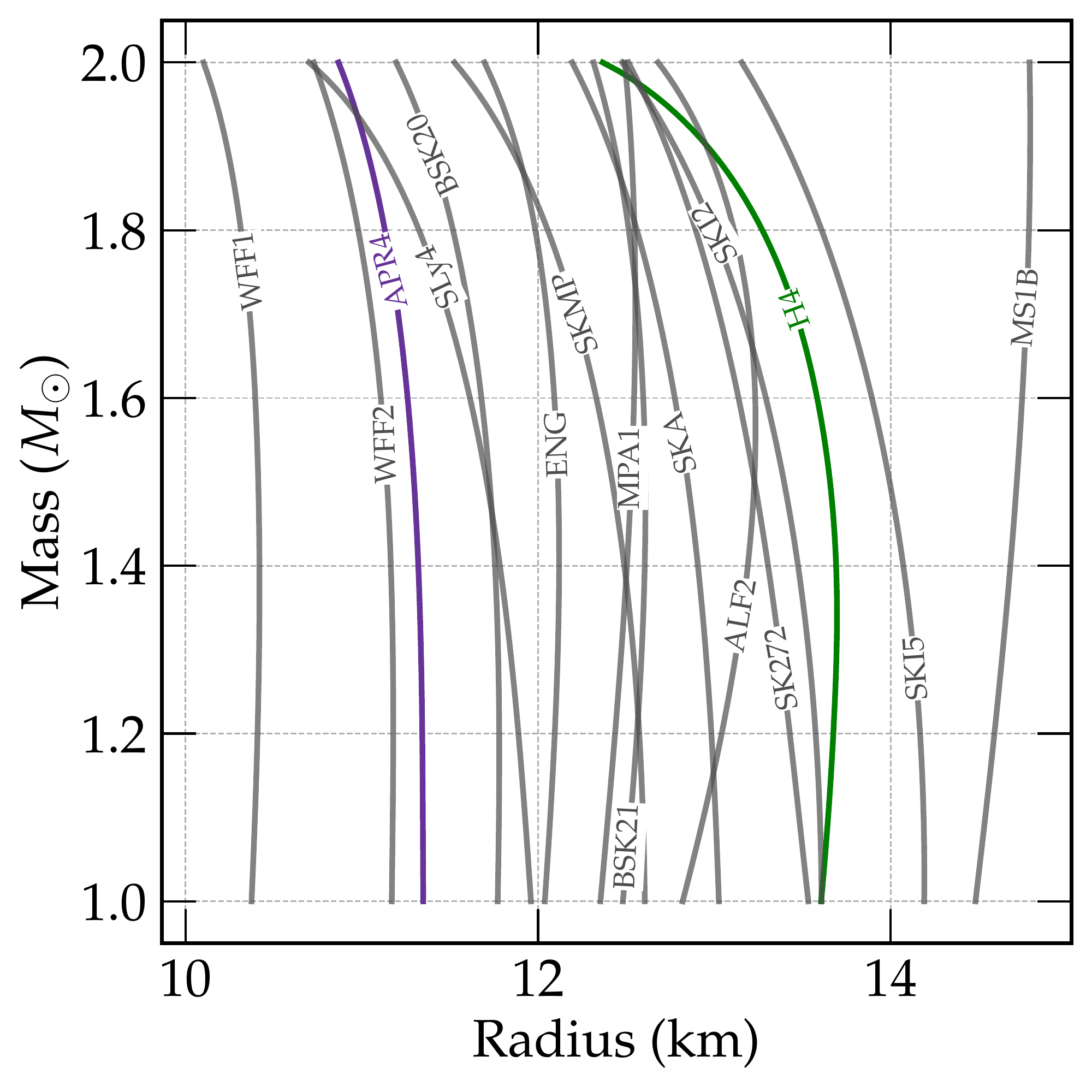}}
  \caption{The mass--radius curves for samples of EoSs, over a mass range of $1-2\,M_{\odot}$.  We choose the APR4 (purple) and H4 (green) EoSs for this study, representing more and less compact NSs, respectively. Produced using LALSimulation \cite{NS_lalsim,lalsuite}.}
  \label{Fig.M-R}
\end{figure}

Here, we consider the Cowling approximation \cite{cowling_1941,Cowling2}, where the perturbations of the metric are neglected, and only fluid perturbations are taken into account \cite{asteroseismology}. Each nuclear-matter EoS generates a unique relation between the mass and radius of NSs \cite{eos_2013}. We need that mass--radius relation to use the empirical relations given above for the frequency and damping time. But the EoSs of NSs are currently not precisely known. However, there are constraints on the EoS from EM and GW observations \cite{doi:10.1146/annurev-nucl-102419-124827}. The mass--radius relations for nonrotating NS models with various EoSs \cite{NS_lalsim,osti_eos} for masses between $1-2\,M_{\odot}$ are shown in Fig.~\ref{Fig.M-R}. In this work, we choose two different EoSs, namely, APR4 \cite{apr4} and H4 \cite{h4}, which are considered as representatives for the classes of soft (more compact) and hard (less compact) models, respectively.

\begin{figure}[t!] 
  \centering
    \scalebox{0.25}{\includegraphics{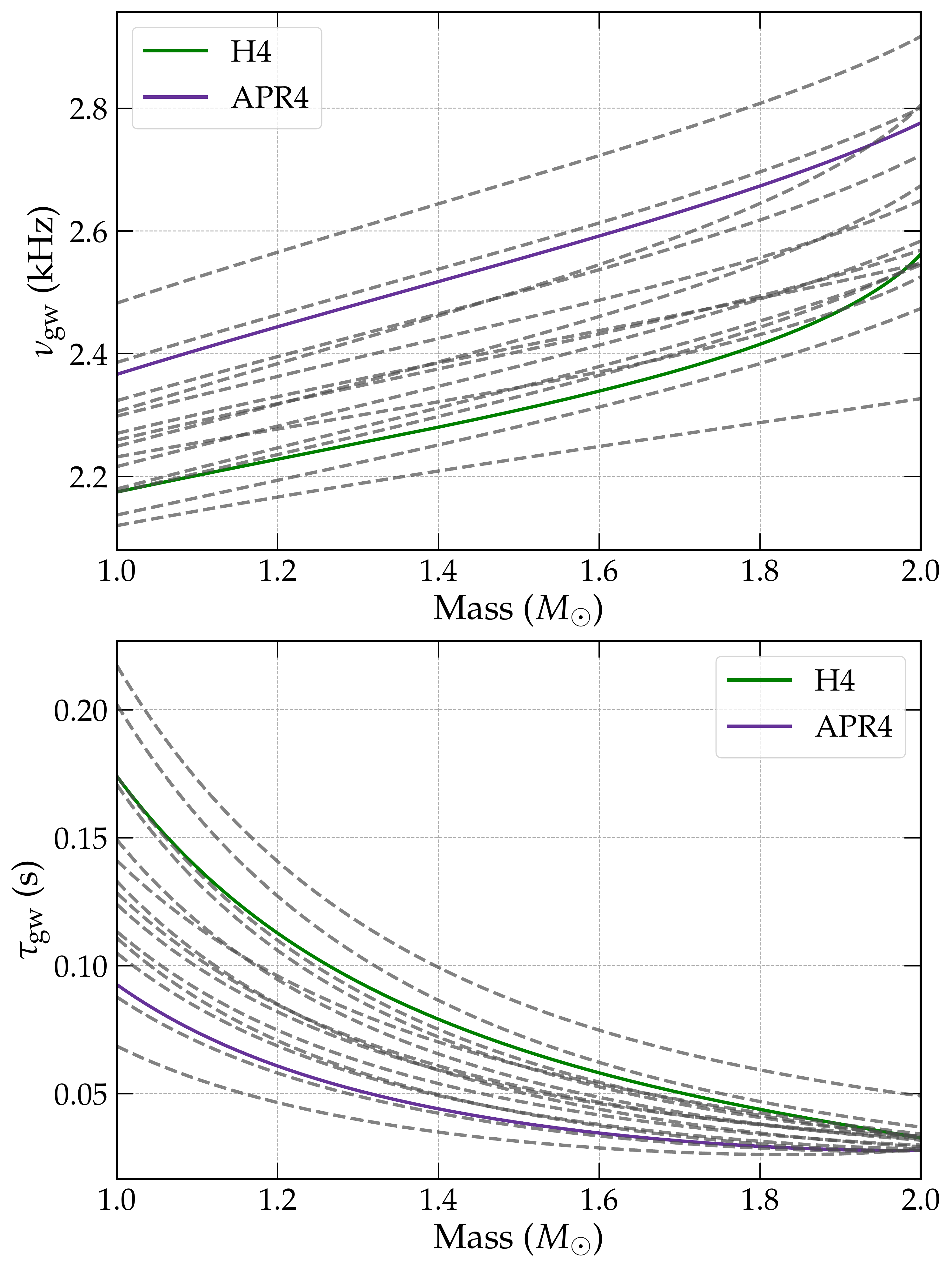}}
  \caption{GW frequency (top panel) and damping time (bottom panel) as a function of NS mass according to Eqs.~(\ref{f_freq}) and (\ref{f_tau}) for the cases of EoSs refereed  to in Fig.~\ref{Fig.M-R}. We choose two EoSs representing soft (APR4) and hard (H4) in this work. The gray dotted lines represent the other EoSs from  Fig.~\ref{Fig.M-R}.}
  \label{Fig.2dist}
\end{figure}

For the different  cases of EoSs  from Fig.~\ref{Fig.M-R}, we compute the relation given in Eqs.~(\ref{f_freq}) and (\ref{f_tau}) to obtain the dependency of GW signal properties (frequency and damping time) as a function of NS mass as shown in Fig.~\ref{Fig.2dist}. For the typical masses of NSs between $1-2\,M_{\odot}$, we obtain corresponding $f$-mode frequencies between $2.2-2.8$\,kHz \cite{Jaiswal_2020}. As the mass of the NS increases, the $f$-mode frequency increases. Also, we can infer that for the softer EoS the frequency is systematically higher than that for the harder EoS. One can speculate that a confident detection of GWs from an NS glitch with a sufficiently accurate frequency estimation will contain information about the EoS, which is degenerate with NS mass but not in the entirety of the parameter space. In Fig.~\ref{Fig.2dist}, we also show the distribution of the damping time $\tau_\mathrm{gw}$. The damping times decrease as the mass of the NS increases, and it should be noted that for the softer EoS the damping time drops off faster than for the harder EoS. 

A comparison study for the GW frequency from a $f$-mode oscillation as a function of mean mass density from previous literature is given in the Appendix. The Cowling approximation used in Eq.~(\ref{f_freq}) results in an overestimate of the frequency by about $30\%$. However, using it in the following analysis can be considered a conservative choice as GW detector sensitivities in the kHz regime fall off as the frequency increases.

\section{Analysis Overview} 
\label{sec:searches}

\subsection{Search method and data}
We consider an all-sky search for generic short-duration GW transients using the coherent WaveBurst (cWB) pipeline \cite{DRAGO2021100678}. cWB is a morphology-independent algorithm for the detection and reconstruction of GW transients. It is based on maximum likelihood-ratio statistics applied to excess power above the detector noise in the multiresolution time--frequency representation of GW strain data \cite{cwb_Klimenko,DRAGO2021100678,cwb_page}.
We use the same version of cWB as was used for the LIGO--Virgo--KAGRA collaboration's third observing run (O3) high-frequency search for generic transients \cite{allskyo3}, with the same settings.

\begin{figure}[t!] 
    \scalebox{0.25}{\includegraphics{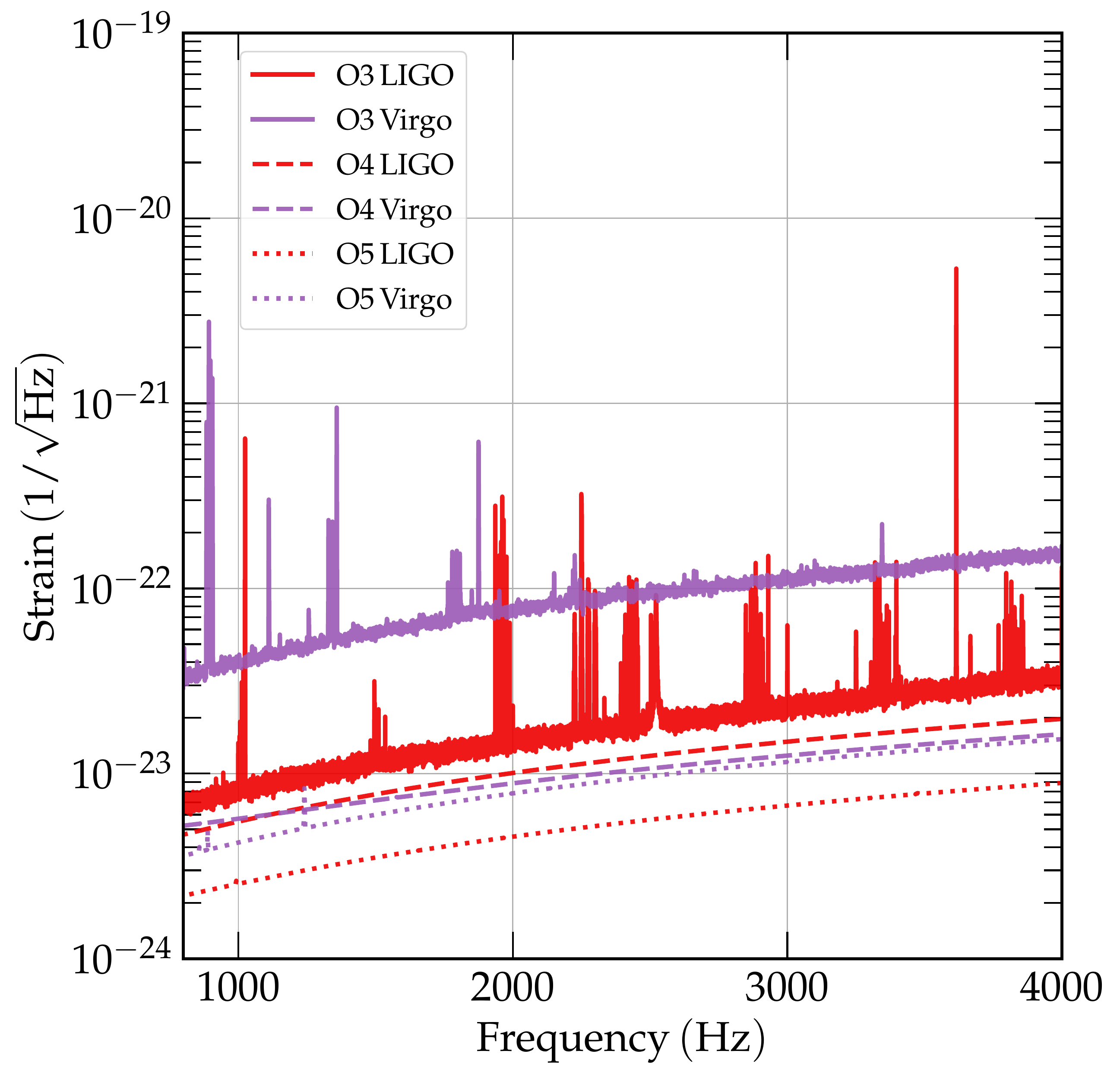}}
\caption{Noise amplitude spectral densities of the LIGO Livingston and Virgo detectors for O3 (measured) as well as predictions for O4 and O5 in the high-frequency range.}
\label{fig:PSD}
\end{figure}

As discussed in Sec.~\ref{sec:signal} and also in \cite{asteroseismology, Andersson1998TowardsGW}, GWs from $f$-mode oscillations in glitching pulsars will occur in the high-frequency range ($2.2-2.8$\,kHz) of ground-based detectors. Therefore, we restrict our analysis to the frequency range of $1-4$\,kHz. We have analyzed the publicly available O3 data, which extended from April 1, 2019, to March 1, 2020 \cite{allskyo3,GWOSC}. For these O3 results, we consider only the Hanford–Livingston (HL) network since Virgo has a significant sensitivity imbalance for frequencies higher than 1\,kHz (almost a factor 5). For the near future prospects of detecting GWs from NS glitches in the fourth (O4) and fifth (O5) observing runs, we have generated Gaussian noise based on the expected spectral sensitivities for the three-detector network with both LIGOs and Virgo \cite{ObservingScenario}. Figure~\ref{fig:PSD} shows the sensitivities of the detectors in terms of measured noise amplitude spectral densities from O3 and the expected curves for O4 and O5 \cite{psd_dcc}.  

\subsubsection{Background generation}

The background is generated by time shifting the detector's  data with respect to the reference detector. For O3, we have used the data from the two LIGO detectors (Livingston and Hanford) and produced over 500 years of time-shifted background with approximately 200 days of available coincident observing time. For O4 and O5, we have simulated 16.85 days of data for the LIGO and Virgo detectors by assuming Gaussian noise which follows the PSD for the corresponding detector and observing run. From this data, we have produced 23 years of background by a time shifting of LIGO Hanford and Virgo by keeping LIGO Livingston as a reference.
For O3, the most significant trigger had a false-alarm rate of about one event in 0.3 years which is well within the expected background rate (assuming the detectors' glitches follow a Poisson distribution, the significance is 0.5 sigma). The central frequency of this trigger was 2.1\,kHz \cite{allskyo3}.

\subsection{Injection set}
To compute the detection sensitivity of this high-frequency all-sky search setup with a given detector configuration, we perform an injection study of adding simulated damped sinusoid waveforms to the  detector data for O3, O4, and O5 runs. 

In this work, we have used a different injection set in terms of the extrinsic parameters of the signals as compared to the results presented in \cite{allskyo3}. In \cite{allskyo3}, the distribution of the simulated sources in the sky was uniform.
Here, we have used a distribution in sky directions that is uniform over the galactic disk based on the Miyamoto-Nagai galactic disk Model \cite{MN_disk,MNGD_2}. 
Also, in \cite{allskyo3}, the inclination angle of all sources was chosen as face-on (optimally oriented), whereas here we sample uniformly over the full range of inclination angles. The injections are distributed in terms of  root-mean-squared amplitude, $h_{\mathrm{rss}} = \sqrt{\int_{-\infty}^{\infty} \left(h^2_+(t) + h^2_{\times}(t) \right) \mathrm{d}t}$ as in \cite{allskyo3} with injected $h_{\mathrm{rss}}$ value given by $(\sqrt{3})^{\mathrm{N}} 5 \times 10^{-23}$ $\mathrm{Hz}^{-1/2}$, for $\mathrm{N}$ ranges from 0 to 8.

\begin{figure}[t!] 
     \centering
        \scalebox{0.25}{\includegraphics{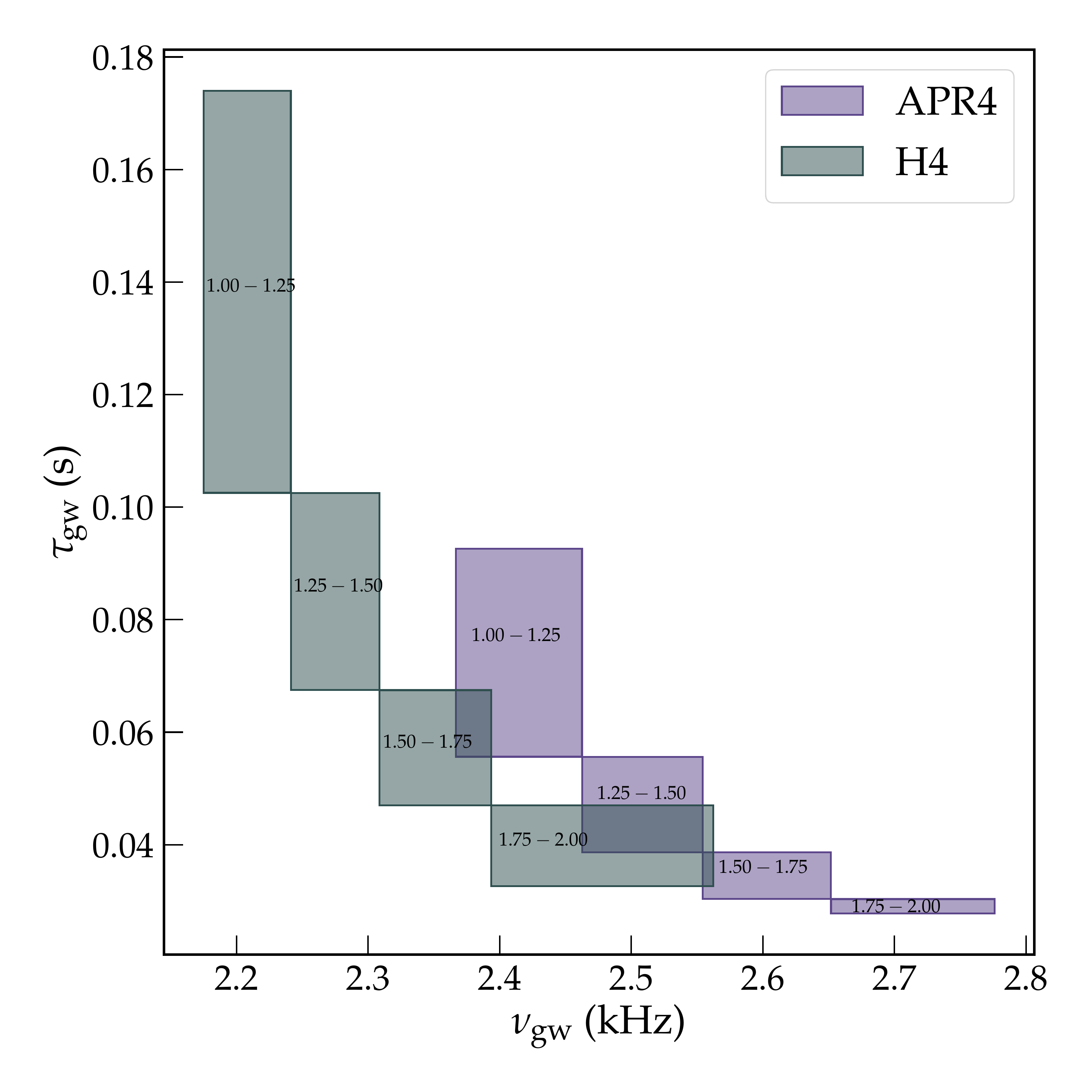}}
    \caption{Distribution of intrinsic parameters (GW frequency and damping time) of injected damped sinusoid waveforms for the four mass bins of the EoS, APR4, and H4. The width and height of each box indicate the spread in frequency and damping time for each injection set.}
        \label{fig:signal_par}
\end{figure}

As discussed in Sec.~\ref{sec:signal}, the intrinsic parameters of the damped sinusoids (frequency and damping time) can be related to the source parameters (mass and EoS of the NS). The mass of the NS is considered from $1-2\,M_{\odot}$ with a bin size of $0.25\,M_{\odot}$. The injection sets are built for each mass bin of the two EoSs considered for this work, which leads to eight injection sets. Figure~\ref{fig:signal_par} shows the distribution of the frequency and damping time of the waveform in the injection set. Each of the eight injection sets considered here is populated with more than 100,000 injections for O3 and 40,000 injections for O4 and O5, providing a precise measurement of the detection efficiency.
 
The amplitude of the incoming signal is a function of distance to the source, spin frequency of the NS, and glitch size. To interpret the results, one can fix any two of the parameters listed above. We fix the distance of the source to that of the Vela pulsar at 287\,pc \cite{vela_dist}. (For clarity, we underline that we do not fix the sky direction to that of Vela, just the distance.) We also fix the spin frequency of the NS to approximately that of Vela ($\nu_{\mathrm{s}} = 11.2$\,Hz) \cite{vela_spin}, and hence, we discuss the results in terms of glitch size.

\subsection{Sensitivity to GW signals during pulsar glitches} 

\begin{figure}[t!] 
     \centering
        \scalebox{0.25}{\includegraphics{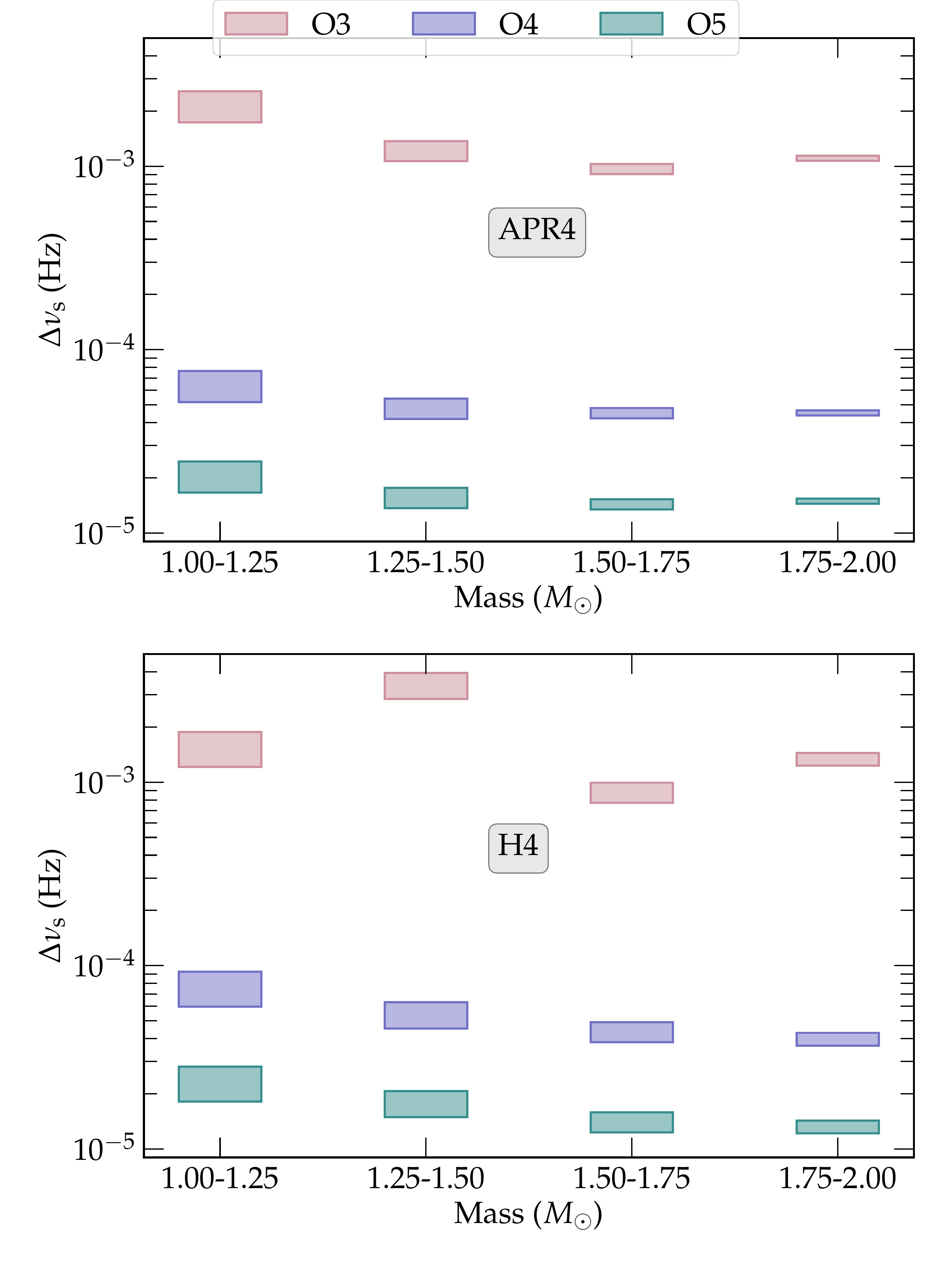}}
    \caption{Sensitivity of the high-frequency all-sky transient search during the O3 run in terms of detectable NS glitch sizes. The corresponding expected sensitivities for O4 and O5 using their predicted noise curves are also shown. Signals are simulated using the spin frequency and distance of the Vela pulsar; other source parameters are drawn from distributions as described in the text with NS models following two different EoSs (APR4 and H4). The sensitivities are shown as separated into four mass bins between $1-2\,M_{\odot}$. The variation in detectable glitch size within each mass bin is indicated by the vertical height of the box for each bin. The glitch size is computed from the minimum $h_{\mathrm{rss}}$ needed for $50\%$ detection efficiency at iFAR$\geq 10$ years. We consider the HL network for O3 results and the HLV network for the O4 and O5 runs.}
        \label{fig:glithc_mag}
\end{figure}

The sensitivity is determined using the value of the quantity $h_{\mathrm{rss}}$ needed to achieve $50\%$ detection efficiency for each mass bin and EoS at an inverse false-alarm rate (iFAR) larger than 10 years. We keep the parameters distance and spin frequency fixed to those of the Vela pulsar and interpret the result in terms of glitch size $\Delta\nu_{\mathrm{s}}$ using Eq.~(\ref{eq_h0}). Here, the peak amplitude, $h_{0}$, is computed from $h_{\mathrm{rss}}$ numerically.
Figure~\ref{fig:glithc_mag} reports the limit on detectable glitch size as a function of mass and EoS for the O3 run, as well as projected detectable glitch sizes for the future O4 and O5 sensitivities. 

In \cite{allskyo3}, the detectable glitch size for optimally oriented sources (uniformly distributed in all sky directions) was greater than $10^{-4}$\,Hz. Under the uniform galactic source distribution, for O3, we find that we would have needed a glitch size larger than $\approx10^{-3}$\,Hz to confidently detect $50\%$ of events. This difference arises mainly from loosening the condition of optimal orientation. For O4, we see around an order of magnitude improvement for the detectable glitch size across the mass bins for both cases of EoS as compared to O3. For O5, this is around two orders of magnitude improvement in detectable glitch size. These improvements are attributed both to improvements in each detector but also the inclusion of Virgo, which allows injections to be recovered from a wider portion of the sky.

The assumption of Gaussian noise (which we make for future observing runs) is not too far from reality for the high-frequency range of the detectors. However, nonstationary lines~\cite{Covas:2018vzm,LIGO:2021ppb} are present in real data.
These lines lead to an anomalously inadequate sensitivity visible in the results (Fig.~\ref{fig:glithc_mag}) for the H4 EoS mass bin $1.25-1.5\,M_{\odot}$.
Following up on this, we found that this reduced sensitivity correlates with a population of lines occurring between $2.2-2.3$\,kHz, which reduces the sensitivity of the detectors for signals falling in this frequency range, leading to higher glitch sizes needed for detections in this mass bin.

\begin{figure}[t!] 
     \centering
        \scalebox{0.25}{\includegraphics{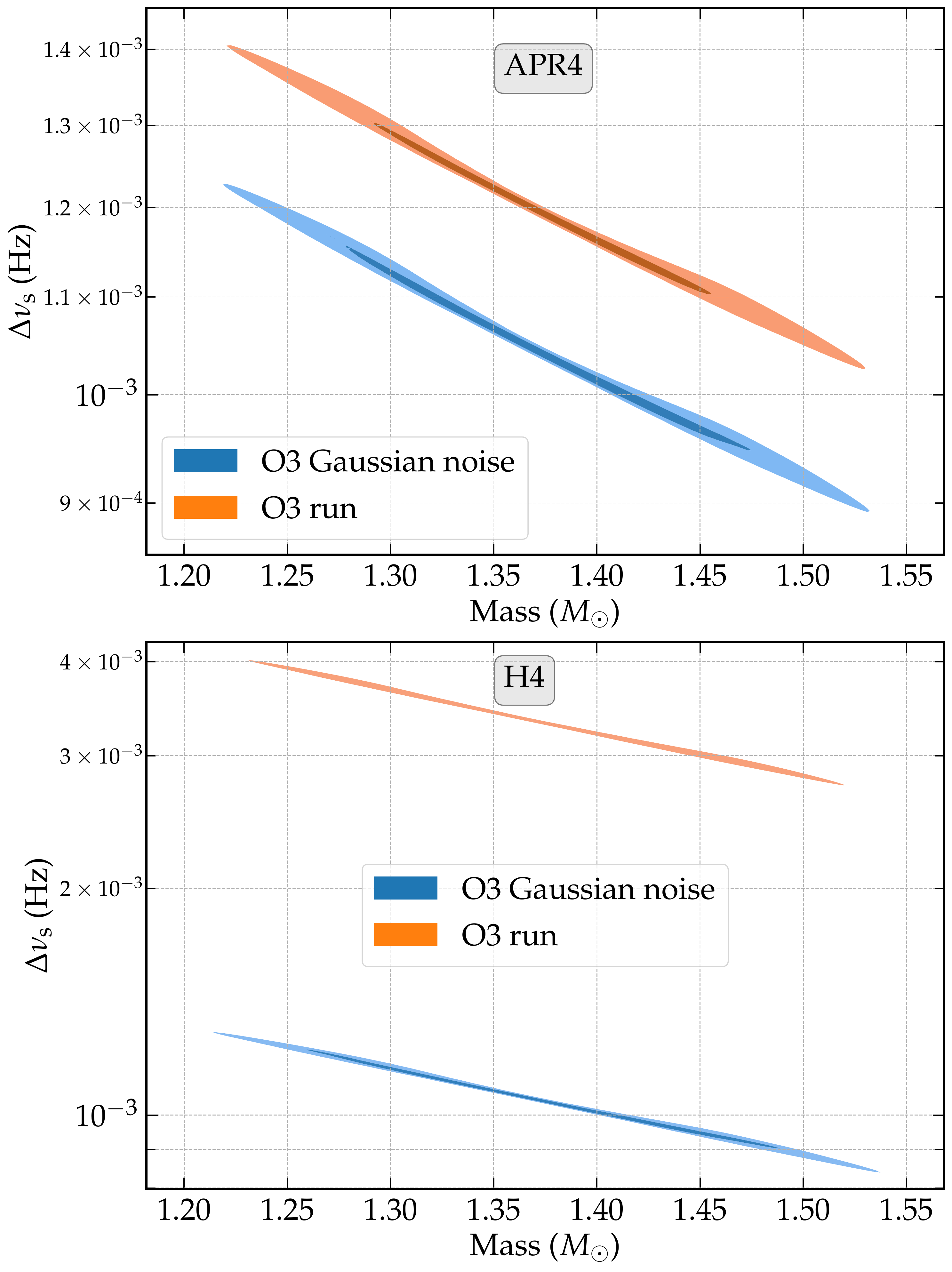}}
    \caption{Difference in detectable glitch sizes obtained from the real O3 data compared with simulated Gaussian noise at O3 detector sensitivity. The injected waveforms are for APR4 and H4 EoS with a mass range between $1.25-1.50\,M_{\odot}$. It can be seen that the nonstationary lines at $2.2-2.3$\,kHz hinder the detectability of H4 EoS signals in this mass range, whereas for APR4 signals which do not fall near the nonstationary lines, detectability is similar in real data and Gaussian noise.}
\label{fig:glithc_diff}
\end{figure}

We conducted an additional study to quantify this hypothesis that indeed the noise happening between  2.2--2.3\,kHz during the O3 run causes this reduction in sensitivity.
For this, we generated simulated data with Gaussian noise based on O3 noise spectral density \cite{psd_dcc} and computed the detectable glitch size at iFAR higher than 10 years.
We found a factor of 3.35 improvement with Gaussian noise as compared to real O3 noise as shown in Fig.~\ref{fig:glithc_diff}.
If we take this into account, this outlier mass bin in H4 EoS with worse sensitivity can be explained. 
This also outlines the fact that the main challenge for the practical implementation of this analysis in future observing runs will be the mitigation of wandering lines in the high-frequency part of the parameter space.

\subsection{Reconstruction of signal's frequency} 
In this section, we briefly report the capability of our search algorithm to reconstruct the injected signal's central frequency. The central frequency of both the injected and reconstructed signal is defined as the energy averaged frequency of all the pixels representing the signal. The robust reconstruction of the frequency is necessary for the identification of the trigger as a possible candidate coming from glitching NS. This can allow for further follow-up with dedicated parameter estimation and search for EM counterparts. 

\begin{figure}[t!] 
     \centering
        \scalebox{0.25}{\includegraphics{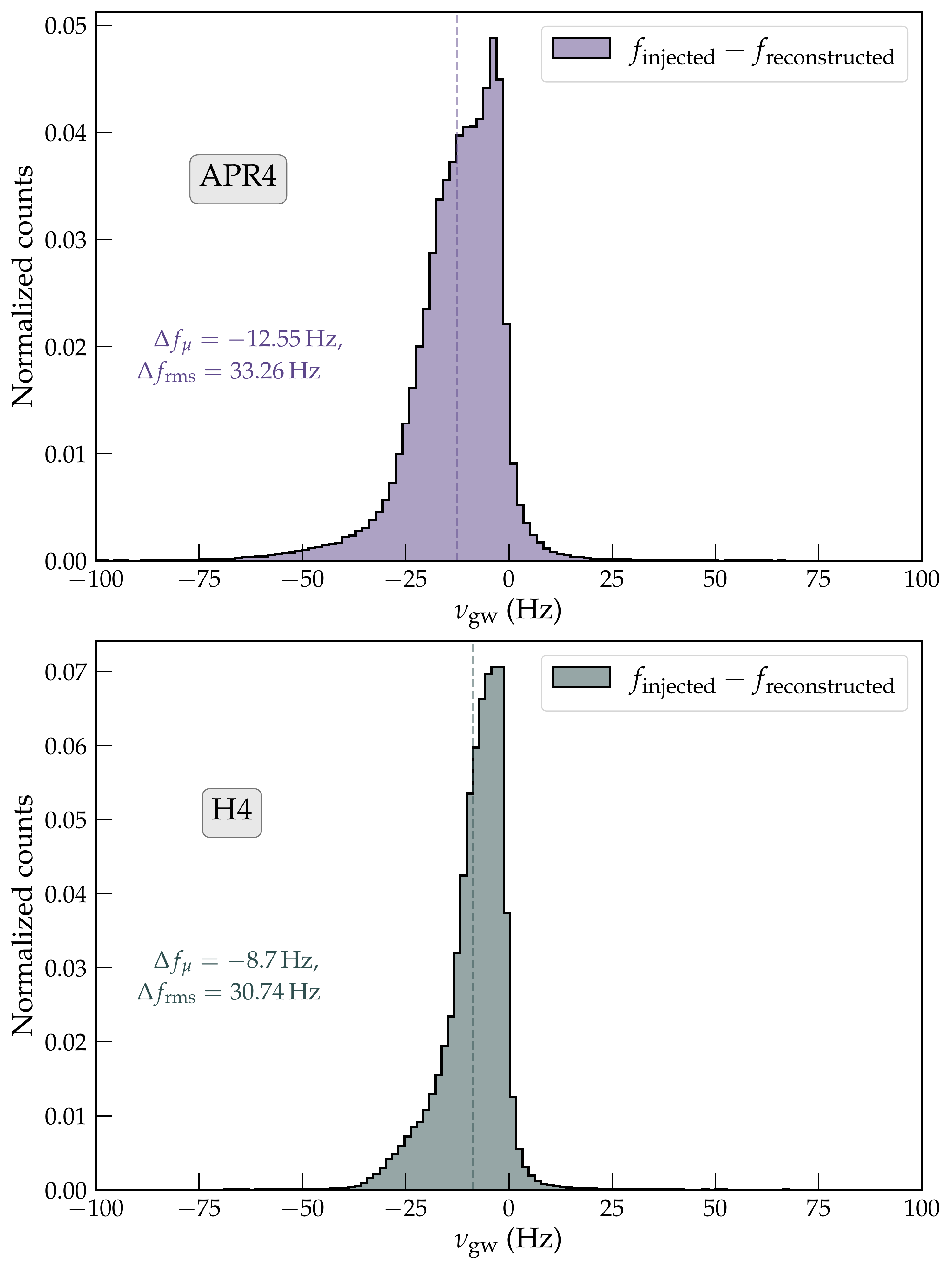}}
    \caption{Difference between the injected and reconstructed central frequencies ($\Delta f$) are plotted for O3 data for the mass bin $1.25-1.50\,M_{\odot}$ and the two EoSs considered. We show the mean $\Delta f_{\mu}$ and rms $\Delta f_{\mathrm{rms}}$ of the distribution. Given the frequencies are over 2\,kHz, the mean and rms are less than $1 \%$. }
\label{fig:rec_inj}
\end{figure}

The reconstruction of the signal's central frequency is shown in Fig.~\ref{fig:rec_inj}. The example shown here is for the O3 data for a mass bin of $1.25-1.50\,M_{\odot}$ for the two EoSs considered in this paper. For all the other injection sets, the results are similar. We show the difference between the injected and recovered frequencies and find that the mean is reconstructed with slightly higher frequencies with a bias of less than $0.5 \%$ and a root-mean-square (rms) of $1 \%$.

\section{Prospects for localizing glitching NS from GW detection}
\label{sec:localization}

Since we are looking to unveil a population of nearby NSs which may not yet have been observed in the EM spectrum, sky localization of the sources of GW detections will play a key role in enabling the follow-up of these events by various ground and space-based telescopes. Even if the GW signal is associated in time with an EM transient, we would still need sky direction information to unambiguously associate the events. In this section, we provide a study on the sky localization capabilities of different networks of GW detectors. 

Sky localization with a network of GW detectors mainly relies upon the time delay measurement between various detectors. Given a pair of detectors, the time of arrival and the amplitude will localize the signal to a ring in the sky \cite{sky_loc,Fairhurst_2010}.
If we have three detectors and hence two pairs, we can localize a source to a much smaller region around the intersection of the two circles in the sky corresponding to each pair of detectors.
Longer baselines between detectors and a higher number of detectors lead to better localization \cite{skyloc_2018}.

In the case of cWB, the likelihood is computed and is maximized over sky directions, which is very sensitive to the time delays, antenna pattern response, and polarization of incoming GWs.
The reconstructed sky direction statistic is a function of the likelihood.
Further discussion about the properties of the sky statistics of cWB can be found in Sec.~III of \cite{cwb_Klimenko}. 

\begin{figure}[t!] 
     \centering
        \scalebox{0.25}{\includegraphics{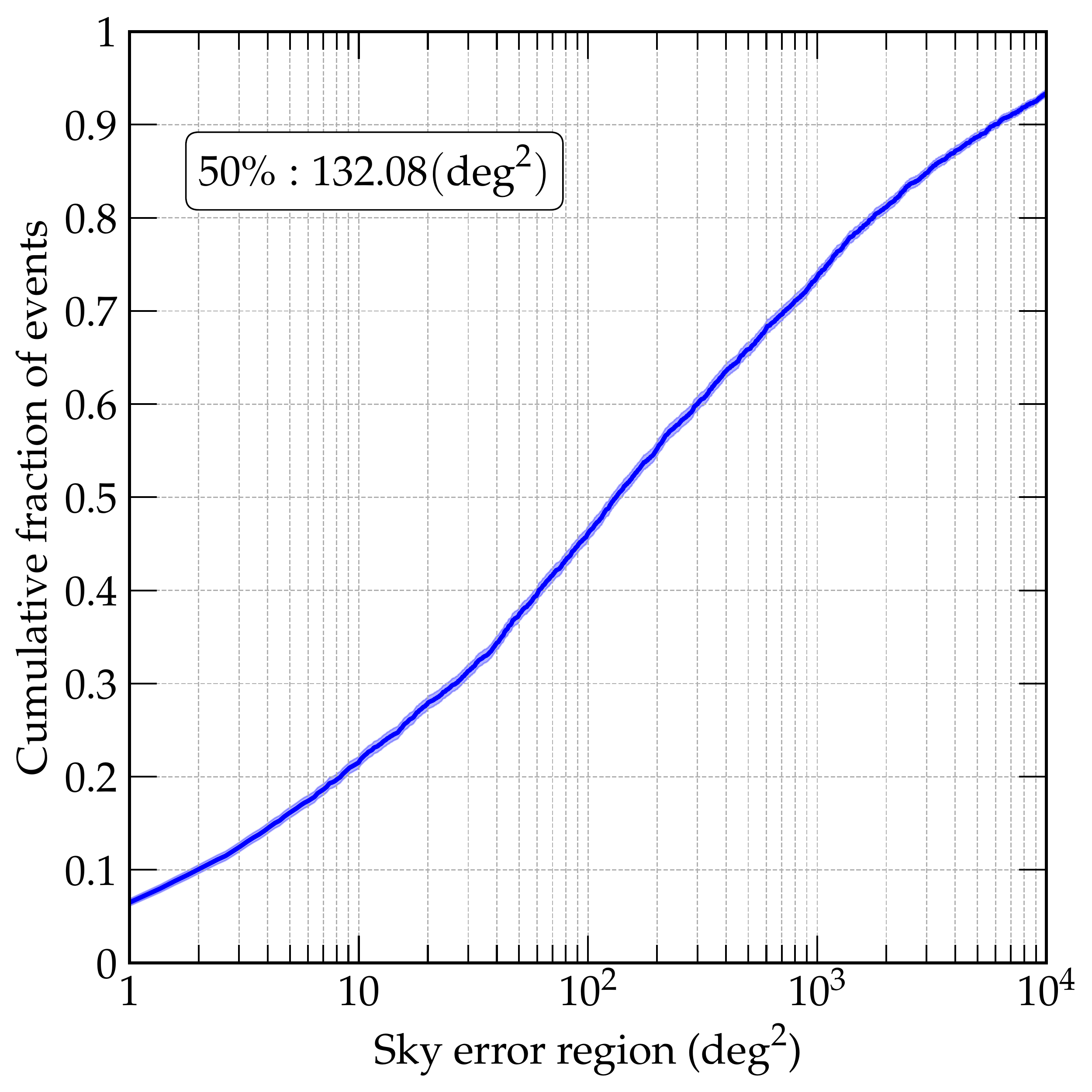}}
    \caption{Sky error region at $1\sigma$  credible interval (SER) for LHVIK network with O5 sensitivity. The blue curve shows the localization area for a cumulative fraction of events with the $1\sigma$ deviation shown in the shaded region. The analysis is illustrated for the injection set with APR4 EoS and mass range from $1.25-1.5\,M_{\odot}$ at injected $\Delta \nu_{s}$ of $1.849 \times 10^{-5}$\,Hz. $50\%$ of injected events are recovered at 132.08~ $\mathrm{deg}^2$.}
        \label{fig:search_area}
\end{figure}

\begin{figure}[t!] 
     \centering
        \scalebox{0.25}{\includegraphics{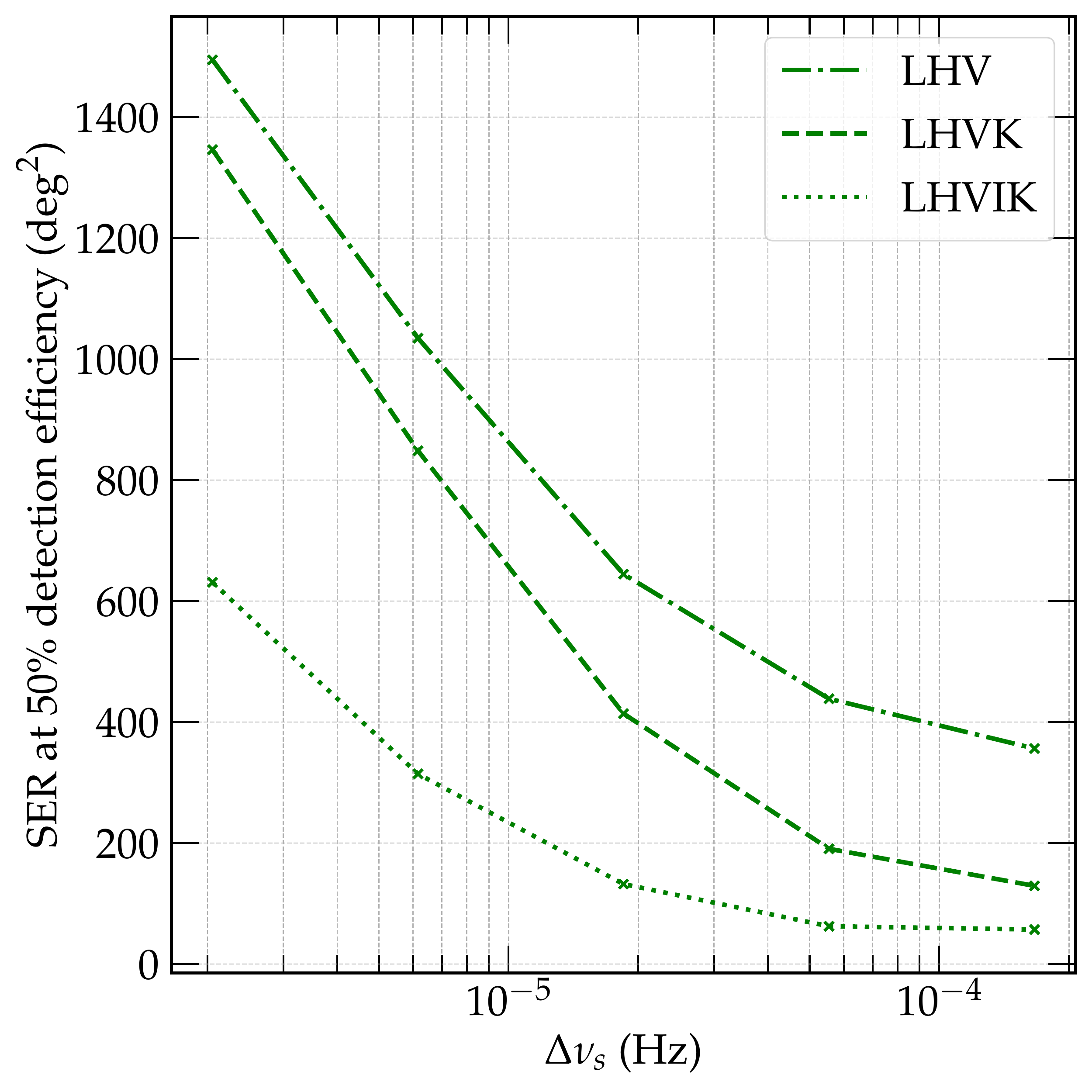}}
    \caption{Sky error region at $1\sigma$ credible interval (SER) as a function of NS glitch size assuming O5 sensitivity for the LHVIK, LHVK, and LHV networks. 
    Specifically, this is the localization region size at $1\sigma$ confidence achieved by 50\% of the events at a given glitch size.
    Here, we used the APR4 EoS and NS masses between $1.25-1.5\,M_{\odot}$ to generate simulated signals.}
        \label{fig:sky_area}
\end{figure}

Here we study the prospects for localizing GW transients from a NS glitch considering the example of APR4 EoS for masses between $1.25-1.5\,M_{\odot}$ and using simulated Gaussian noise for the O5 run. Although we have considered only one injection set for this study, for the other injection sets the results are not expected to change significantly. We inject the NS glitch waveforms in the simulated data corresponding to the spin frequency and distance of the Vela pulsar with five glitch size values spaced between $10^{-6}$\,Hz and $10^{-3}$\,Hz. For each considered glitch size, the sources are uniformly distributed in sky direction and source orientation, the same as in \cite{allskyo3}, to get the maximum efficiency. The metric we use here to determine typical sky localization performance is the sky error region at a 1$\sigma$ credible interval for 50\% of the detected events\cite{Klimenko2001}. 
We consider the following present and future detectors: LIGO-Hanford (H), LIGO-Livingston (L), Virgo (V), KAGRA (K), and LIGO-India (I), combining them in three different networks: LHV, LHVK, and LHVIK. In Fig.~\ref{fig:search_area}, we show the cumulative histogram for the LHVIK network for glitch size $10^{-5}$\,Hz as a function of the sky error region. We can see that only 10\% of the detected events are localized better than $2 \ \mathrm{deg}^{2}$, whereas 50\% of the events are localized with the sky error region better than 132 $ \mathrm{deg}^{2}$. With the choice of the sky error region for 50\% of the events, we explore various network configurations and glitch sizes in Fig.~\ref{fig:sky_area}.     

As expected, a drastic improvement can be seen as the number of detectors in the network increases and also an improvement as the glitch size grows (since the signal-to-noise ratio grows with it). For a scenario discussed before of a $10^{-5}$\,Hz glitch size with the five detector networks, we get the sky localization region of around 132~$\mathrm{deg}^{2}$ at $1\sigma$ uncertainty for 50\% of events, which might be too large for many EM telescopes to efficiently follow-up the full sky area, but there can be a few cases where the sky localization can be as good as a few $\mathrm{deg}^2$. Compared with CBC localization areas in O1--O3 \cite{gwtc1,gwtc2,gwtc3}, this level of localization can still provide an opportunity to potentially find an EM counterpart to a transient burst GW detection from a glitching pulsar.   

\section{Discussion}
\label{sec:discussions}

In this work, we have updated the all-sky upper limits from the LIGO--Virgo O3 run for short-duration GW signals from $f$ modes triggered by glitches in NSs without electromagnetic counterparts by using more realistic distributions in the extrinsic parameters (i.e., sky direction and orientation of the source) of the simulated signals used for sensitivity estimation.
We have also investigated how line artifacts in the O3 data affect sensitivity in certain frequency bands and hence source mass ranges, which sheds light on the practical challenges which we face for such high-frequency short-duration searches. Further, we give the prospects for the detection and localization of such short-duration GWs from NS glitches for the upcoming fourth (O4) and fifth (O5) observing runs of the current generation of ground-based detectors.
By fixing the reference pulsar as Vela (in terms of distance and spin frequency), we found that the detectable glitch size will be around $10^{-4}$\,Hz for O4 and $10^{-5}$\,Hz for O5. Glitch sizes of $10^{-5}$\,Hz have been observed by radio telescopes before \cite{315_glitches, vela_spin, parkes_radio}. Further, it has been shown that observed pulsar glitches form two different populations when it comes to glitch size \cite{vela_type,Fuentes:2017bjx,Arumugam:2022ugq}. These distributions are conventionally called \textit{normal} or \textit{Crab-like} for the smaller glitches and \textit{Vela-like} for the larger (mean at around $10^{-4.4}$\,Hz) glitches. Thus, for O4 and O5, there can be a more realistic chance to observe a nearby glitching pulsar if the glitch comes from the population of \textit{Vela-like} glitches. We have also studied the localization capability of this type of GW search, finding that with a five detector network during O5 for the detectable glitch sizes of $10^{-5}$\,Hz the EM follow-up can be challenging, as the sky error region for $50\%$ events at $1\sigma$ is about 132~$\mathrm{deg}^{2}$.
However, this localization can still be useful for future wide-scope telescopes like CHIME, SKA, etc. \cite{chime, ska}.
It could also be sufficient to associate the GW event with the galactic disk.

The proposed third-generation GW observatories like Einstein Telescope \cite{et_science}, Cosmic Explorer \cite{cs_science}, and NEMO \cite{Nemo} will provide much higher sensitivities at kHz ranges, ideal for observing GW signals from glitching pulsars. We leave it for future studies to quantify the sensitivity of third generation detectors, where the analysis methods and search configurations will need to be very different. 

\begin{acknowledgments}
We thank Leigh Smith for comments on an early version of this draft. The document has been given LIGO DCC number P2200190. D.L. acknowledges support from Swiss National Science Foundation (SNSF) Grant No.~200020-182047. S.T. is supported by Swiss National Science Foundation (SNSF) Ambizione Grant No.~PZ00P2-202204. M.D. acknowledges the support from the 
Amaldi Research Center funded by the MIUR program 
`Dipartimento di Eccellenza' (CUP:B81I18001170001) and 
the Sapienza School for Advanced Studies (SSAS).

D.K. is supported by the Spanish Ministerio de Ciencia, Innovaci\'on y Universidades (ref.~BEAGAL 18/00148)
and cofinanced by the Universitat de les Illes Balears,
and acknowledges support by European Union FEDER funds, 
the Spanish Ministerio de Ciencia e Innovaci\'on,
and Spanish Agencia Estatal de Investigaci\'on Grants No.
PID2019-106416GB-I00/MCIN/AEI/10.13039/501100011033,
RED2018-102661-T,
RED2018-102573-E,
the European Union NextGenerationEU funds (PRTR-C17.I1);
the Comunitat Aut\`onoma de les Illes Balears through the Direcci{\'o} General de Pol{\'i}tica Universitària i Recerca with funds from the Tourist Stay Tax Law ITS 2017-006 (PRD2018/24, PDR2020/11);
the Conselleria de Fons Europeus, Universitat i Cultura del Govern de les Illes Balears;
the Generalitat Valenciana (PROMETEO/2019/071);
and EU COST Actions CA18108 and CA17137.

This research has made use of data obtained from the 
Gravitational Wave Open Science Center \cite{gwsc}, a service of LIGO Laboratory, 
the LIGO Scientific Collaboration and the Virgo Collaboration. The authors are grateful for computational resources provided by the LIGO Lab (CIT) and supported by National Science Foundation Grants No.~PHY-0757058 and No.~PHY-0823459. This material is based upon work supported by NSF's LIGO Laboratory which is a major facility fully funded by the National Science Foundation.

\end{acknowledgments}

\appendix

\renewcommand{\thesubsection}{\Alph{subsection}}
\section*{APPENDIX: COMPARISON OF \textit{f}-MODE FREQUENCIES}
\label{app:fmode_freq}

Here, we are looking for the frequency of a GW signal emitted by an $f$-mode oscillation during a glitch that is related to the mean density of the NS \cite{asteroseismology, Andersson1998TowardsGW, Kokkotas:2001ze}. 
These relations are found from the solution of the nonradial perturbations equation of a nonrotating star in general relativity (GR) or using the Cowling approximation \cite{rotating_NS,gr_damptime,Cowling2,cow_gr90,cow_gr97,Chandrasekhar1,Chandrasekhar2,astroses2004}.

We compare our injected signal frequency with numerical relativity simulations solving the Einstein equations for dynamical spacetimes in full GR present in literature. The empirical fit for fundamental mode frequency $\nu$ in full GR with dynamical spacetime for a nonrotating star is given as
\begin{equation}\label{fmode_freq} 
    \nu[\mathrm{kHz}] = k_l+\mu_l\left(\frac{\bar{M}}{\bar{R}^3}\right)^{1/2} \,.
\end{equation}

\begin{figure}[H] 
  \centering
    \scalebox{0.25}{\includegraphics{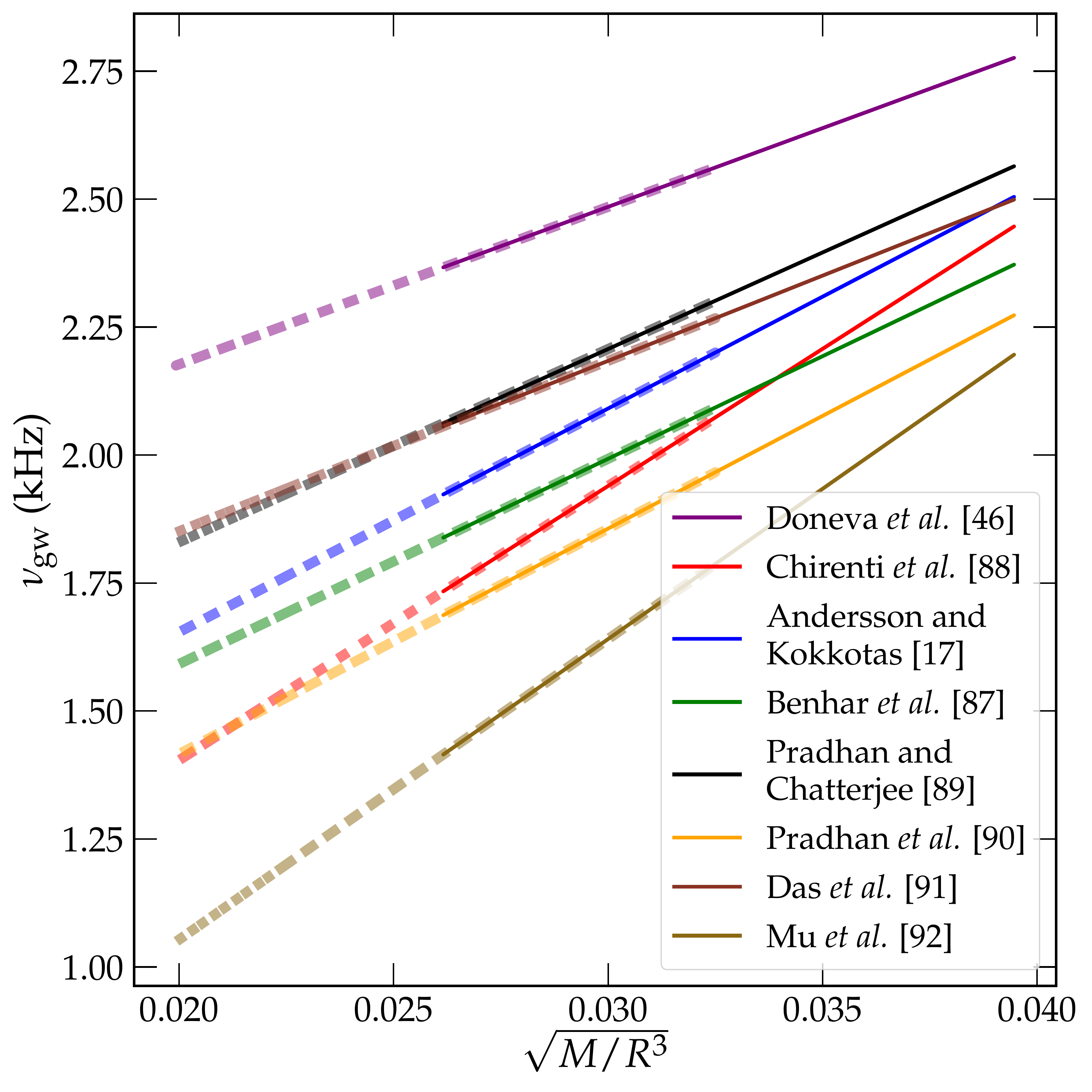}}
  \caption{Distribution of GW frequency as a function of NS mean density. Each curve shows the $f$-mode oscillation frequency derived using Eq.~(\ref{fmode_freq}) from the literature mentioned in Table~\ref{tab:f_freq}. The solid (dotted) line shows the APR4 (H4) EoS for the NS mass ranges from $1-2\,M_{\odot}$.}
  \label{Fig.freq_den}
\end{figure}

Table~\ref{tab:f_freq} show the coefficients $k_{l}$ and $\mu_{l}$ of Eq.~(\ref{fmode_freq}) for $l=2$ given in different literature \cite{asteroseismology,Andersson1998TowardsGW,astroses2004,fmode_gr2,Pradhan2021,Pradhan2022,Das_2021,Mu_2022}. The corresponding GW frequencies as a function of average density for the EoS, APR4, and H4 are shown in Fig.~\ref{Fig.freq_den}. It clearly shows the frequency is overestimated with the Cowling approximation in the nonrotating limit \cite{asteroseismology}. 

\begin{table}[ht]
\centering
\begin{tabular}{c c c}
\hline  \hline
Reference & $k_{2}$ & $\mu_{2}$ \\ [1ex] 
 \hline \hline
Doneva  \textit{et al.} \cite{asteroseismology} & 1.562 & 1.151\\
Andersson and Kokkotas \cite{Andersson1998TowardsGW} & 0.78 & 1.635\\
Benhar \textit{et al.} \cite{astroses2004} & 0.76 & 1.5\\
Chirenti \textit{et al.} \cite{fmode_gr2} & 0.332 & 2.005\\
Pradhan and Chatterjee \cite{Pradhan2021} & 1.075 & 1.412\\
Das \textit{et al.} \cite{Das_2021} & 1.185 & 1.246\\
Mu \textit{et al.} \cite{Mu_2022} & --0.121 & 2.197\\
Pradhan \textit{et al.} \cite{Pradhan2022} & 0.535 & 1.648\\[1ex] 
\end{tabular}
\caption{Coefficients $k_{l}$ and $\nu_{l}$ of Eq.~(\ref{fmode_freq}) ($l=2$) for $f$-mode frequency from different literature.}
\label{tab:f_freq} 
\end{table}

\clearpage

\bibliography{ref}

\end{document}